\documentstyle[preprint,aps,epsf]{revtex}

\newcommand{\be}{\begin{equation}}
\newcommand{\ba}{\begin{eqnarray}}
\newcommand{\ee}{\end{equation}}
\newcommand{\ea}{\end{eqnarray}}
\newcommand{\nn}{\nonumber}

\begin{document}
\draft
\preprint{PITHA 96/18 (Revised)}
\title{\bf Baryon Magnetic Moments and Proton Spin :\\
A Model with Collective Quark Rotation}

\author{Massimo Casu\footnote{email: casu@physik.rwth-aachen.de} and 
L.~M.~Sehgal\footnote{email: sehgal@physik.rwth-aachen.de}}
\address{Institute for Theoretical Physics (E), RWTH Aachen\\
D-52074 Aachen, Germany}

\maketitle

%%%%%%%%%%%%%%%%%%%%%%%%%%%%%%%%%%%%%%%%%%%%%%%%%%%%%%%%%%%%%%%%%%%%%

\begin{abstract}
We analyse the baryon magnetic moments in a model that relates them to
the parton spins $\Delta u$, $\Delta d$, $\Delta s$, and includes a 
contribution from orbital angular momentum. The specific assumption is
the existence of a 3-quark correlation (such as a flux string) that 
rotates with angular momentum $\langle L_z \rangle$ around the proton
spin axis. A fit to the baryon magnetic moments, constrained by the measured
values of the axial vector coupling constants $a^{(3)} = F + D$, $a^{(8)} = 
3F - D$, yields $\langle S_z \rangle = 0.08 \pm 0.13$, $\langle L_z \rangle
= 0.39 \pm 0.09$, where the error is a theoretical estimate. A second fit,
under slightly different assumptions, gives $\langle L_z \rangle = 0.37 \pm 
0.09$, with no constraint on $\langle S_z \rangle$. The model provides a consistent 
description of axial vector 
couplings, magnetic moments and the quark polarization $\langle S_z \rangle$ 
measured in deep inelastic scattering. The fits suggest that a significant 
part of the angular momentum of the proton may reside in a collective 
rotation of the constituent quarks.
\end{abstract} 

%%%%%%%%%%%%%%%%%%%%%%%%%%%%%%%%%%%%%%%%%%%%%%%%%%%%%%%%%%%%%%%%%%%%%

\newpage

\section{Introduction}
The question of the angular momentum composition of the proton, first raised
in the context of the quark parton model in 1974 \cite{Sehgal}, has developed into
a burning issue, following experiments on polarized deep inelastic scattering,
and progress in the theoretical under\-standing of QCD. Within the quark parton
model, the contribution of polarized quarks and antiquarks to the spin of a
polarized proton ($J_z = 1/2$) is \cite{Sehgal}
\ba
\langle S_z \rangle &=& \frac{1}{2} ( \Delta u + \Delta d + \Delta s ) \equiv
\frac{1}{2} \Delta\Sigma \nn\\
\mbox{with} \quad \Delta\Sigma &=& (3F-D) + \delta_{\mathrm{EJ}} 
\ea
Here $\Delta q$ is the net polarization of quarks of flavour $q$, 
$\Delta q = \int dx [ \{ q_+(x) - q_-(x) \} + \{ \bar{q}_+(x) - \bar{q}_-(x) 
\} ]$, $F$ and $D$ are the axial vector  coupling constants of $\beta$-decay 
( $F = 0.462 \pm 0.01$, $D = 0.794 \pm 0.01$; Ref. \cite{Song} ), and 
$\delta_{\mathrm{EJ}}$ is the ``defect'' in the Ellis-Jaffe sum rule
\cite{Ellis}
\be
\delta_{\mathrm{EJ}} = \int g_1^p(x) dx - \left ( \frac{1}{2}F -
\frac{1}{18}D \right )
\ee

In QCD, the expression for $\Delta\Sigma$ is modified by perturbative gluon
corrections \cite{Kodaira} and by a contribution from the gluon anomaly in
the singlet axial vector current \cite{Efremov}, and reads
\be
\Delta\Sigma = (3F-D) + \delta_{\mathrm{EJ}}(Q^2) + \delta_{\mathrm{anomaly}}
\ee
where, to lowest order in $\alpha_s/ \pi$,
\be
\delta_{\mathrm{EJ}}(Q^2) = \left ( 1 - \frac{\alpha_s(Q^2)}{\pi} \right )^{-1}
\int g_1^p(x,Q^2) dx - \left ( \frac{1}{2}F - \frac{1}{18}D \right ) 
\ee
\be
\delta_{\mathrm{anomaly}} = n_f \frac{\alpha_s}{2\pi} \Delta G
\ee
Here $\Delta G$ is the net gluon polarization, $\Delta G = \int dx [ 
G_+(x) - G_-(x) ]$, and $n_f = 3$ is the number of light quark flavours.
A number of authors \cite{Jaffe} have analysed the data \cite{SMC} on the
structure functions $g_1^{p,n}$, and have reached the conclusion that, barring 
a large correction from the anomalous term $\delta_{\mathrm{anomaly}}$, 
$\Delta\Sigma$ lies in the interval
\be
\Delta\Sigma \simeq ( 0.1 \ldots 0.3 )
\ee
Thus the polarization of the quarks and antiquarks accounts for only $10-30\%$ of 
the spin of the proton, a typical solution for the spin decomposition being
$\Delta u = 0.83 \pm 0.03$, $\Delta d = -0.43 \pm 0.03$, $\Delta s = -0.10 \pm
0.03$ \cite{Karliner}. 

%%%%%%%%%%%%%%%%%%%%%%%%%%%%%%%%%%%%%%%%%%%%%%%%%%%%%%%%%%%%%%%%%%%%%

\section{The baryon magnetic moments}
In Ref. \cite{Sehgal}, a tentative attempt was made to relate the nucleon
magnetic moments to the spin structure of the proton, encoded in the
parameters $\Delta u$, $\Delta d$, $\Delta s$. This idea has recently been
generalized to the full baryon octet in two papers \cite{Karl,Bartelski} that
have investigated the following ansatz for the magnetic moments :
\ba
\mu(p) &=& \mu_u \delta u + \mu_d \delta d + \mu_s \delta s \nn\\
\mu(n) &=& \mu_u \delta d + \mu_d \delta u + \mu_s \delta s \nn\\
\mu(\Sigma^+) &=& \mu_u \delta u + \mu_d \delta s + \mu_s \delta d \nn\\
\mu(\Sigma^-) &=& \mu_u \delta s + \mu_d \delta u + \mu_s \delta d \\
\mu(\Xi^-) &=& \mu_u \delta s + \mu_d \delta d + \mu_s \delta u \nn\\
\mu(\Xi^0) &=& \mu_u \delta d + \mu_d \delta s + \mu_s \delta u \nn\\
\mu(\Lambda^0) &=& \frac{1}{6} ( \delta u + 4 \delta d + \delta s ) ( \mu_u 
+ \mu_d ) + \frac{1}{6} ( 4 \delta u - 2 \delta d + 4 \delta s ) \mu_s \nn
\ea
The baryon magnetic moments are linear combinations of $\delta u$,
$\delta d$, $\delta s$, defined by $\delta q = \int dx [ \{ q_+(x) - q_-(x)
\} - \{ \bar{q}_+(x) - \bar{q}_-(x) \} ]$, which differs from $\Delta q$ in
the sign of the antiquark contribution. We consider two hypotheses for the
relation between $\delta q$ and $\Delta q$:
\begin{itemize}
\item[A.] Antiquarks in a polarized baryon reside entirely in a cloud of
spin-zero mesons. In this case, antiquarks have no net polarisation, i.e.
$\bar{q}_+ - \bar{q}_- = 0$, so that $\delta q = \Delta q$. Models of this
type have been discussed, for instance, by Cheng and Li \cite{Cheng}.
\item[B.] Antiquarks in a polarized baryon are generated entirely by the
pertubative splitting of gluons $g \to q\bar{q}$. In such a case, it is
reasonable to expect $\bar{u}_+ - \bar{u}_- \approx \bar{d}_+ - \bar{d}_-
\approx \bar{s}_+ - \bar{s}_- \approx s_+ - s_-$. The corresponding relation 
between $\delta q$ and $\Delta q$ is $\delta u = \Delta u - \Delta s$, 
$\delta d = \Delta d - \Delta s$, $\delta s = 0$ (see, e.g. Ref. 
\cite{Bartelski}).
\end{itemize}
Below, we give the results of fits to the baryon magnetic moments based on each 
of the above two hypotheses.
 
\underline{\bf Fit A.} Assumption A implies that Eqs.(7) may be rewritten
with $\delta q$ replaced by $\Delta q$. Such an approximation was considered
by Karl \cite{Karl}, who concluded that the data could be fitted with values 
of $\Delta u$, $\Delta d$, $\Delta s$ similar to those deduced from
polarized deep inelastic scattering, and that the fit was superior to
that given by the conventional quark model characterised by $\Delta u =
4/3$, $\Delta d = -1/3$, $\Delta s = 0$. Our own results for model A are
shown in Table 1. As in Ref. \cite{Karl}, each magnetic moment was assigned
a theoretical uncertainty of $\pm 0.1 \mu_N$. This (arbitrary) choice
ensures that the various magnetic moments have approximately equal weight
and that the fits have a $\chi^2$ of about one unit per degree of freedom.
The conventional quark model result is given under the appellation ``Model 0''. Note
that this model necessarily implies a nucleon axial vector coupling $G_A
\equiv a^{(3)} = F+D = \Delta u - \Delta d = 5/3$, in conflict with the
measured value 1.26. Notice also that the fit deviates markedly from the
expectation $\mu_u = -2\mu_d$. By contrast, the column labelled ``Model AI''
gives the result of a fit to Eqs.(7) in which $\Delta u$ and $\Delta d$ are
constrained to give the correct value of $G_A$, i.e. $G_A = 1.26$. Additionally, 
we take $\mu_u = -2\mu_d$ and
$\mu_s = 3/5\mu_d$ (the latter assumption agrees with the fitted value in
Ref. \cite{Karl}, and also with the usual constituent quark model estimate
$m_d/m_s = 0.6$). It is convenient to rewrite $\Delta u$, $\Delta d$,
$\Delta s$ as 
\ba
\Delta u &=& \frac{2}{3} S_z + \frac{1}{2} G_A + \frac{1}{6} a^{(8)} \nn\\
\Delta d &=& \frac{2}{3} S_z - \frac{1}{2} G_A + \frac{1}{6} a^{(8)} \\
\Delta s &=& \frac{2}{3} S_z - \frac{1}{3} a^{(8)} \nn
\ea
so that the magnetic moments in Eq. (7) can be treated as functions of three
parameters $\mu_u$, $S_z = \frac{1}{2} ( \Delta u + \Delta d + \Delta s )$
and $a^{(8)} = \Delta u + \Delta d - 2 \Delta s$. The results of the fit are
\ba
\mu_u &=& 2.39 \pm 0.06 \nn\\
S_z &=& 0.14 \pm 0.12 \qquad\qquad\qquad \mbox{(Model AI)}\\
a^{(8)} &=& 0.85 \pm 0.06 \nn
\ea 
For the central value of $\mu_u$, the allowed domain of the parameters $S_z$ 
and $a^{(8)}$ is shown in Fig.1 (ellipse labelled $L_z = 0$). While the value 
of $S_z$ is in good agreement with the determinations from high energy 
scattering, there is a clear discrepancy between the value of $a^{(8)}$ 
obtained from the fit and its experimental value $a^{(8)} = 3F-D \approx 0.60$.
\newline

\underline{\bf Fit B.} We now repeat the analysis of the magnetic moments
using the ansatz B. Written in terms of $\Delta q$, Eqs.(7) now involve only
the combinations $a^{(3)}=\Delta u - \Delta d = G_A$ and $a^{(8)}=\Delta u +
\Delta d  - 2\Delta s$, and are independent of the combination $a^{(0)}=\Delta u
+ \Delta d + \Delta s = 2S_z$. Accordingly, the fit,\bigskip ~using $G_A=1.26$ as
input, determines only the two parameters\newline
\bigskip
\parbox{11cm}{\begin{eqnarray*}\mu_u &=& 2.40 \pm 0.06 \\
a^{(8)} &=& 0.82 \pm 0.05 \end{eqnarray*}} \hfill
\parbox{5cm}{\begin{eqnarray}\!\!\!\!\!\!\!\!\!\!\!\!\!\!\!\!\!\!\!\!\!\!\!
\!\!\!\!\!\!\!\!\!\!\!\!\!\!\!\!\!\!\!\!\!\!\!\!\!\!\!\!\!\!\!\!\!\!\!\!\!
\!\!\!\!\!\!\!\!\!\!\!\!\!\!\!\!\!\!\!\!\!\mbox{(Model BI)}\end{eqnarray}}
no constraint being obtained on $S_z$. The allowed domain of these two
parameters is shown in Fig.2 by the ellipse labelled $L_z = 0$. The
value of $a^{(8)}$ in Eq.(10) is very similar to the value in Fit A, Eq.(9).
In both cases, however, the value of $a^{(8)}$ deviates significantly from
the value measured in hyperon decay.

%%%%%%%%%%%%%%%%%%%%%%%%%%%%%%%%%%%%%%%%%%%%%%%%%%%%%%%%%%%%%%%%%%%%%

\section{The rotating proton}
In an attempt to resolve the above discrepancy, we have constructed a model 
containing orbital angular momentum. The
total angular momentum of a polarized proton can be resolved as $J_z = S_z + L_z +
\Delta G = \frac{1}{2}$. We consider here the effects of an orbital angular
momentum $\langle L_z \rangle$ associated with the motion of three
constituent quarks in the baryon. As pointed out in \cite{Sehgal}, such
orbital motion will produce a correction to the magnetic moments, dependent
on the way in which the angular momentum $\langle L_z \rangle$ is shared
between the constituents. Our central hypothesis is that the quarks in a
baryon are held together by a flux string in a ``Mercedes-star''
configuration. In the plane transverse to the proton spin axis, the quarks
will tend to be situated at the corners of an equilateral triangle (Fig.3).
Let us imagine that this correlated 3-quark structure rotates collectively
around the z-axis, with total orbital angular momentum $\langle L_z \rangle$.
For a baryon containing constituents $q_1$, $q_2$, $q_3$ with masses $m_1$,
$m_2$, $m_3$, the orbital angular momentum carried by the quark $q_i$ is
$[ m_i / ( m_1 + m_2 + m_3 ) ] \langle L_z \rangle$ (we assume rotation
about the geometrical centre of the triangle, thereby maintaining SU(3)
symmetry in the baryon spatial wave function). With this simple ansatz, we
obtain the following corrections to the seven baryon magnetic moments listed
in Eq.(7) :
\ba
\mu(p) &=& \ldots + \left [ 2 \mu_u \left ( \frac{1}{3} \right ) + \mu_d
\left ( \frac{1}{3} \right ) \right ] \langle L_z \rangle \nn\\
\mu(n) &=& \ldots + \left [ \mu_u \left ( \frac{1}{3} \right ) + 2 \mu_d
\left ( \frac{1}{3} \right ) \right ] \langle L_z \rangle \nn\\
\mu(\Sigma^+) &=& \ldots + \left [ 2 \mu_u \left ( \frac{\lambda}{1+2\lambda} 
\right ) + \mu_s \left ( \frac{1}{1+2\lambda} \right ) \right ] 
\langle L_z \rangle \nn\\
\mu(\Sigma^-) &=& \ldots + \left [ 2 \mu_d \left ( \frac{\lambda}{1+2\lambda} 
\right ) + \mu_s \left ( \frac{1}{1+2\lambda} \right ) \right ] 
\langle L_z \rangle \\
\mu(\Xi^-) &=& \ldots + \left [ \mu_d \left ( \frac{\lambda}{2+\lambda} \right ) 
+2\mu_s \left ( \frac{1}{2+\lambda} \right ) \right ] \langle L_z \rangle \nn\\
\mu(\Xi^0) &=& \ldots + \left [ \mu_u \left ( \frac{\lambda}{2+\lambda} \right )
+2\mu_s \left ( \frac{1}{2+\lambda} \right ) \right ] \langle L_z \rangle \nn\\
\mu(\Lambda^0) &=& \ldots + \left [ \mu_u \left ( \frac{\lambda}{1+2\lambda} 
\right ) + \mu_d \left ( \frac{\lambda}{1+2\lambda} \right )  + \mu_s \left (
\frac{1}{1+2\lambda} \right ) \right ] \langle L_z \rangle \nn
\ea
where $\lambda = m_d/m_s$ is taken to be $0.6$, and the dots 
``$\ldots$'' represent the spin contribution given in Eq.(7).

We have fitted the seven magnetic moments under the same assumptions employed 
in models A and B (namely, $a^{(3)} = \Delta u - \Delta d = 1.26$, $\mu_u = 
-2\mu_d$, $\mu_s = \frac{3}{5}\mu_d$), using $\langle L_z \rangle$ as an 
additional parameter. In a first variation of model A, the parameter 
$\langle L_z \rangle$ was fixed such that $\langle L_z \rangle + \langle S_z 
\rangle = \frac{1}{2}$. This represents the extreme hypothesis that the 
``missing'' angular momentum of the proton is precisely accounted for by the 
orbital angular momentum of the correlated structure depicted in Fig.3. This 
model then contains the same free parameters as Model AI, namely $\mu_u$, $S_z$ 
and $a^{(8)}$. A fit to the magnetic moments (see Table 1) yields
\ba
\mu_u &=& 2.17 \pm 0.09 \nn\\
S_z &=& 0.11 \pm 0.14 \qquad\qquad\qquad \mbox{(Model AII)}\\
a^{(8)} &=& 0.60 \pm 0.10 \nn
\ea 
The quality of the fit is essentially the same as in Model AI, but there is 
a dramatic improvement in the value of $a^{(8)}$, the result of the fit 
coinciding with the measured value. This improvement is evident from Fig.1, 
which shows that with the inclusion of $L_z$ there is a convergence of the
data on magnetic moments, axial vector couplings and polarized deep
inelastic scattering. Within the framework of ansatz A, we can also
consider $\langle S_z \rangle$ and $\langle L_z \rangle$ as independent
free parameters, using the experimental value of $a^{(8)}$ as input. A
three-parameter fit to the magnetic moments then yields
\ba
\mu_u &=& 2.17 \pm 0.08 \nn\\
\langle S_z \rangle &=& 0.08 \pm 0.13 \qquad\qquad\qquad \mbox{(Model AIII)}\\
\langle L_z \rangle &=& 0.39 \pm 0.09 \nn
\ea

If the effects of orbital angular momentum given by Eqs.(10) are incorporated
into model B, we obtain the results indicated in columns BII and BIII in Table
2. A three-parameter fit in terms of $\mu_u$, $L_z$ and  $a^{(8)}$ yields
\ba
\mu_u &=& 2.10 \pm 0.19 \nn\\
\langle L_z \rangle &=& 0.54 \pm 0.37 \qquad\qquad\qquad \mbox{(Model BII)}\\
\langle L_z \rangle &=& 0.49 \pm 0.23 \nn
\ea
On the other hand, if $a^{(8)} = 0.6$ is used as input,\bigskip ~we find\newline
\bigskip
\parbox{11cm}{\begin{eqnarray*}\mu_u &=& 2.19 \pm 0.08 \\ 
L_z &=& 0.37 \pm 0.09 \end{eqnarray*}} \hfill 
\parbox{5cm}{\begin{eqnarray}\!\!\!\!\!\!\!\!\!\!\!\!\!\!\!\!\!\!\!\!\!\!\!
\!\!\!\!\!\!\!\!\!\!\!\!\!\!\!\!\!\!\!\!\!\!\!\!\!\!\!\!\!\!\!\!\!\!\!\!\!
\!\!\!\!\!\!\!\!\!\!\!\!\!\!\!\!\!\!\!\!\!\mbox{(Model BIII)}\end{eqnarray}}
The improved convergence of magnetic moment and axial vector coupling data
in the presence of orbital angular momentum is evident from Fig.2. Also
noteworthy is the similarity in the fitted value of $\langle L_z \rangle$ in
models A and B, Eqs. (13) and (15). It is certainly intriguing that the value
of $\langle L_z \rangle$ derived from fits to the static properties of baryons (magnetic moments
and axial vector couplings) has the correct sign and
approximately the correct magnitude to explain the ``spin deficit'' of the
nucleon revealed by high energy scattering.

%%%%%%%%%%%%%%%%%%%%%%%%%%%%%%%%%%%%%%%%%%%%%%%%%%%%%%%%%%%%%%%%%%%%%

\section{Conclusion}
It would appear from the above that the quark parton model defined by the 
parton spins $\Delta u$, $\Delta d$, $\Delta s$, can provide a consistent 
description of axial vector couplings, baryon magnetic moments and the 
spin structure functions, provided we supplement the spin angular momentum 
with a collective orbital angular momentum as symbolised in Fig.3. The role 
of the rotating flux string in achieving this agreement draws renewed 
attention to flux-string models of the baryon (see e.g. \cite{Kalashnikova} 
and references therein). Such models have been invoked in the past to
explain states in the baryon spectrum (such as the Roper resonance N(1440))
that have not been easy to accomodate in the traditional three-quark picture 
\cite{Cutkosky}. The idea that the nucleon may contain $L \neq 0$
components in its wave function (``configuration mixing'') has also been
entertained before \cite{Glashow}. The possibility of 
rotation as a source of hadron spin has been emphasised by Yang   
\cite{Yang}. The specific structure introduced in the present paper may be 
expected, naively, to produce rotational levels
with energy $E_{\mathrm{rot}} = J(J+1)/(2I)$, where $I$ is the moment of
inertia of the 3-quark correlation. Assuming this structure to consist of
three constituent quarks in close contact, each with radius 0.2$-$0.3 fm
\cite{Bjorken}, the excitation energy is 0.5$-$1.0 GeV. It remains to be
seen whether the spectrum of baryonic levels will show evidence for states
associated with string-like configurations, beyond those that are expected 
from the shell model with three independently moving quarks.
Direct experimental tests for rotating constituents in the nucleon have been
proposed in \cite{Meng}, and some tentative evidence from hadronic 
reactions has been reported \cite{Borus}.  

%%%%%%%%%%%%%%%%%%%%%%%%%%%%%%%%%%%%%%%%%%%%%%%%%%%%%%%%%%%%%%%%%%%%%

\section{Acknowlegdement}
We wish to record our thanks to Dr. O. Biebel for his help in the error
analysis.

%%%%%%%%%%%%%%%%%%%%%%%%%%%%%%%%%%%%%%%%%%%%%%%%%%%%%%%%%%%%%%%%%%%%%

%%%%%%%%%%%%%%%%%%%%%%%%%%%%%%%%%%%%%%%%%%%%%%%%%%%%%%%%%%%%%%%%%%%%%

\newpage

\begin{center}
{\large Figure Captions}
\end{center}

\begin{itemize}
\item[Fig.1.] Fit to baryon magnetic moments in Model A, compared with value
of $a^{(8)}$ from hyperon decay, and $S_z$ from polarized deep inelastic
scattering (bands correspond to $a^{(8)} = 0.60 \pm 0.05$, $S_z = 0.10 \pm
0.05$). The ellipses labelled $L_z=0$ and $L_z \neq 0$ correspond to the
solutions AI and AII in Table 1.
\item[Fig.2.] Fit to baryon magnetic moments in Model B, compared with
value of $a^{(8)}$ from hyperon decay (band corresponds to $a^{(8)} = 0.60 
\pm 0.05$). The ellipses labelled $L_z=0$ and $L_z \neq 0$ correspond to the
solutions BI and BIII in Table 2.
\item[Fig.3.] Flux string connecting three constituent quarks, rotating
collectively around proton spin axis
\end{itemize}

\vspace*{2cm}

\begin{center}
{\large Table Captions}
\end{center}

\begin{itemize}
\item[Table 1.] Fit to baryon magnetic moments in model A. Magnetic
moments are in nucleon magnetons and the $\pm 0.1$ is a fictive theoretical 
error.
\end{itemize}

\begin{itemize}
\item[Table 2.] Fit to baryon magnetic moments in model B. Magnetic
moments are in nucleon magnetons and the $\pm 0.1$ is a fictive theoretical    
error. 
\end{itemize}

%%%%%%%%%%%%%%%%%%%%%%%%%%%%%%%%%%%%%%%%%%%%%%%%%%%%%%%%%%%%%%%%%%%%%

\newpage
\oddsidemargin-1cm

\begin{center}
\begin{tabular}{c|c|c|c|c|c}
\multicolumn{6}{c}{\rule[-5mm]{0mm}{8mm}\bf TABLE 1.}\\ \hline\hline
& magn.& Model 0 & Model AI & Model AII & Model AIII\\
& moments & $S_z=\frac{1}{2}$, $L_z=0$  & $S_z$ free, $L_z=0$ & 
$S_z+L_z=\frac{1}{2}$ & $S_z$, $L_z$ free\\ \hline
$\mu(p)$ & $\begin{array}{cc} 2.79\pm 0.1 \\ \pm 0.00000006\end{array}$ &
2.67 & 2.68 & 2.74 & 2.74\\ \hline
$\mu(n)$ & $\begin{array}{cc} -1.91\pm 0.1 \\ \pm 0.0000005\end{array}$ &
$-$1.92 & $-$1.84 & $-$1.78 & $-$1.79\\ \hline
$\mu(\Sigma^+)$ & $\begin{array}{cc} 2.46\pm 0.1 \\ \pm 0.01\end{array}$ &
2.54 & 2.58 & 2.52 & 2.52\\ \hline
$\mu(\Sigma^-)$ & $\begin{array}{cc} -1.16\pm 0.1 \\ \pm 0.025\end{array}$ &
$-$1.14 & $-$1.21 & $-$1.20 & $-$1.20\\ \hline
$\mu(\Xi^-)$ & $\begin{array}{cc} -0.65\pm 0.1 \\ \pm 0.0025\end{array}$ &
$-$0.48 & $-$0.60 & $-$0.60 & $-$0.60\\ \hline
$\mu(\Xi^0)$ & $\begin{array}{cc} -1.25\pm 0.1 \\ \pm 0.014\end{array}$ &
$-$1.40 & $-$1.34 & $-$1.38 & $-$1.39\\ \hline
$\mu(\Lambda)$ & $\begin{array}{cc} -0.61\pm 0.1 \\ \pm 0.004\end{array}$ &
$-$0.61 & $-$0.60 & $-$0.60 & $-$0.61\\ \hline
Input & & $\begin{array}{ccc} \Delta u = \frac{4}{3} \\ \Delta d =
-\frac{1}{3} \\ \Delta s=0 \end{array}$ & $\begin{array}{ccc} \mu_u=-2\mu_d & \\
\mu_s=\frac{3}{5}\mu_d & \\ G_A=1.26 \end{array}$ & $\begin{array}{ccc}
\mu_u=-2\mu_d & \\ \mu_s=\frac{3}{5}\mu_d & \\ G_A=1.26 \end{array}$  &
$\begin{array}{cccc} \mu_u=-2\mu_d \\ \mu_s=\frac{3}{5}\mu_d & \\ G_A=1.26
\\ a^{(8)}=0.60 \end{array}$ \\ \hline
$\chi^2/DOF$ & & $1.82$ & $1.12$ & $1.105$ & $1.095$ \\ \hline
$\begin{array}{cc} \mbox{fitted} \\ \mbox{param.} \end{array}$ & &
$\begin{array}{ccc} \mu_u=1.75\pm 0.06 \\ \mu_d=-1.01\pm 0.06 \\
\mu_s=-0.61\pm 0.05 \end{array}$ & $\begin{array}{cccc} \mu_u=2.17\pm 0.09 \\
S_z=0.14\pm 0.12 \\ a^{(8)}=0.85\pm 0.06 \\ \mbox{exp: } 0.60 \pm 0.02
\end{array}$
& $\begin{array}{cccc} \mu_u=2.17\pm 0.09 \\ S_z=0.11\pm 0.14 \\
a^{(8)}=0.60 \pm 0.10 \\ \mbox{exp: } 0.60 \pm 0.02\end{array}$ &
$\begin{array}{ccc} \mu_u=2.17\pm 0.08 \\ S_z=0.08\pm 0.13 \\
L_z=0.39 \pm 0.09 \end{array}$ \\ \hline\hline
\end{tabular}   
\end{center}   

%%%%%%%%%%%%%%%%%%%%%%%%%%%%%%%%%%%%%%%%%%%%%%%%%%%%%%%%%%%%%%%%%%%%%

\newpage                   

\begin{center}             
\begin{tabular}{c|c|c|c|c|c}
\multicolumn{6}{c}{\rule[-5mm]{0mm}{8mm}\bf TABLE 2.}\\ \hline\hline                 
& magn.& Model 0 & Model BI & Model BII & Model BIII\\
& moments & $S_z=\frac{1}{2}$ & $S_z$ undetermined &
$S_z$ undetermined & $S_z$ undetermined \\
&& $L_z=0$ & $L_z=0$ & $L_z$ free & $L_z$ free \\ \hline            
$\mu(p)$ & $\begin{array}{cc} 2.79\pm 0.1 \\ \pm 0.00000006\end{array}$ &
2.67 & 2.76 & 2.81 & 2.80 \\ \hline
$\mu(n)$ & $\begin{array}{cc} -1.91\pm 0.1 \\ \pm 0.0000005\end{array}$ &
$-$1.92 & $-$1.78 & $-$1.73 & $-$1.74 \\ \hline
$\mu(\Sigma^+)$ & $\begin{array}{cc} 2.46\pm 0.1 \\ \pm 0.01\end{array}$ &
2.54 & 2.65 & 2.54 & 2.59 \\ \hline
$\mu(\Sigma^-)$ & $\begin{array}{cc} -1.16\pm 0.1 \\ \pm 0.025\end{array}$ &
$-$1.14 & $-$1.09 & $-$1.14 & $-$1.13 \\ \hline
$\mu(\Xi^-)$ & $\begin{array}{cc} -0.65\pm 0.1 \\ \pm 0.0025\end{array}$ &
$-$0.48 & $-$0.49 & $-$0.54 & $-$0.53 \\ \hline
$\mu(\Xi^0)$ & $\begin{array}{cc} -1.25\pm 0.1 \\ \pm 0.014\end{array}$ &
$-$1.40 & $-$1.28 & $-$1.36 & $-$1.33 \\ \hline
$\mu(\Lambda)$ & $\begin{array}{cc} -0.61\pm 0.1 \\ \pm 0.004\end{array}$ &
$-$0.61 & $-$0.52 & $-$0.57 & $-$0.55 \\ \hline
Input & & $\begin{array}{ccc} \Delta u = \frac{4}{3} \\ \Delta d =
-\frac{1}{3} \\ \Delta s=0 \end{array}$ & $\begin{array}{ccc} \mu_u=-2\mu_d
& \\ \mu_s=\frac{3}{5}\mu_d & \\ G_A=1.26 \end{array}$ & $\begin{array}{ccc}
\mu_u=-2\mu_d & \\ \mu_s=\frac{3}{5}\mu_d & \\ G_A=1.26 \end{array}$  &
$\begin{array}{cccc} \mu_u=-2\mu_d \\ \mu_s=\frac{3}{5}\mu_d & \\ G_A=1.26
\\ a^{(8)}=0.60 \end{array}$ \\ \hline
$\chi^2/DOF$ & & $1.82$ & $1.99$ & $1.72$ & $1.43$ \\ \hline
$\begin{array}{cc} \mbox{fitted} \\ \mbox{param.} \end{array}$ & &
$\begin{array}{ccc} \mu_u=1.75\pm 0.06 \\ \mu_d=-1.01\pm 0.06 \\
\mu_s=-0.61\pm 0.05 \end{array}$ & $\begin{array}{ccc} \mu_u=2.40\pm 0.06 \\
a^{(8)}=0.82\pm 0.05 \\ \mbox{exp: } 0.60 \pm 0.02 \end{array}$
& $\begin{array}{cccc} \mu_u=2.10\pm 0.19 \\ L_z=0.54\pm 0.37 \\
a^{(8)}=0.49 \pm 0.23 \\ \mbox{exp: } 0.60 \pm 0.02\end{array}$ &
$\begin{array}{cc} \mu_u=2.19\pm 0.08 \\ L_z=0.37\pm 0.09 \end{array}$ \\ \hline\hline
\end{tabular}
\end{center}

%%%%%%%%%%%%%%%%%%%%%%%%%%%%%%%%%%%%%%%%%%%%%%%%%%%%%%%%%%%%%%%%%%%%%

\newpage
\oddsidemargin0cm

\begin{figure}[b]
\begin{center}
\mbox{\epsfysize 20cm \epsffile{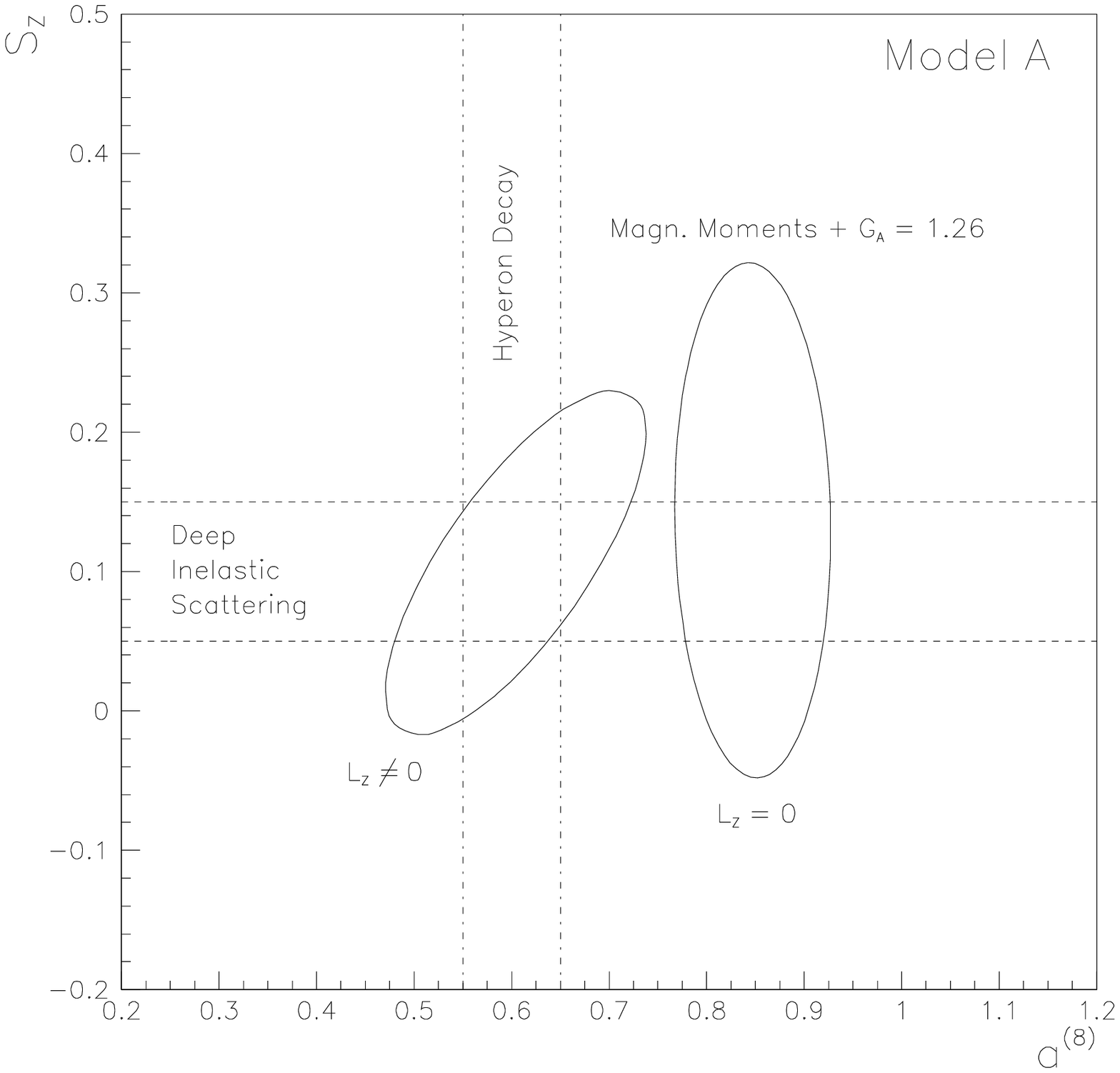}}
\end{center}
\caption{}
\end{figure}

\newpage

\begin{figure}[b]
\begin{center}
\mbox{\epsfysize 20cm \epsffile{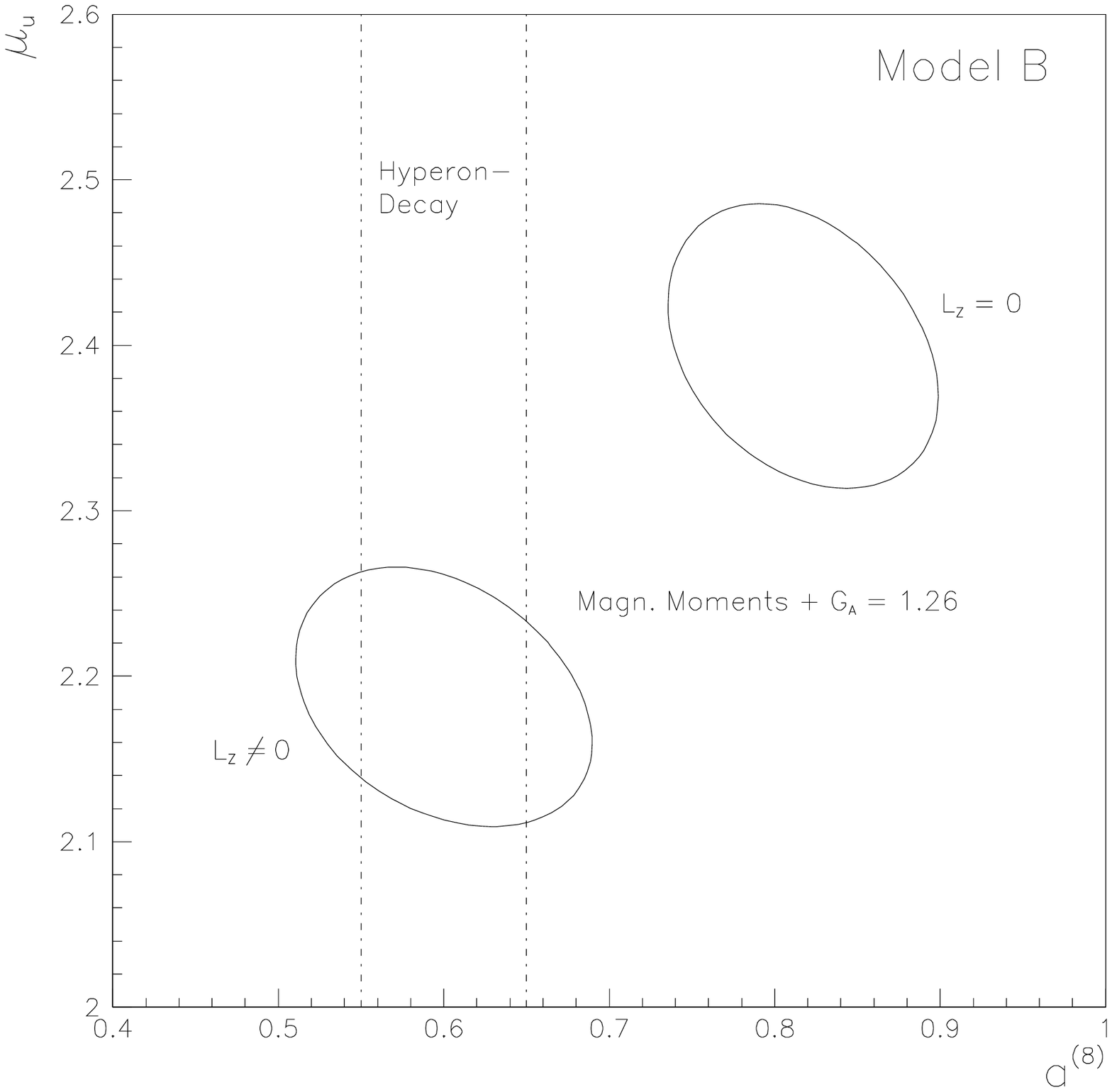}}
\end{center}
\caption{}       
\end{figure}

\newpage
\vspace*{3cm}

\begin{figure}[ht]
\begin{center}
\mbox{\epsfysize 14cm \epsffile{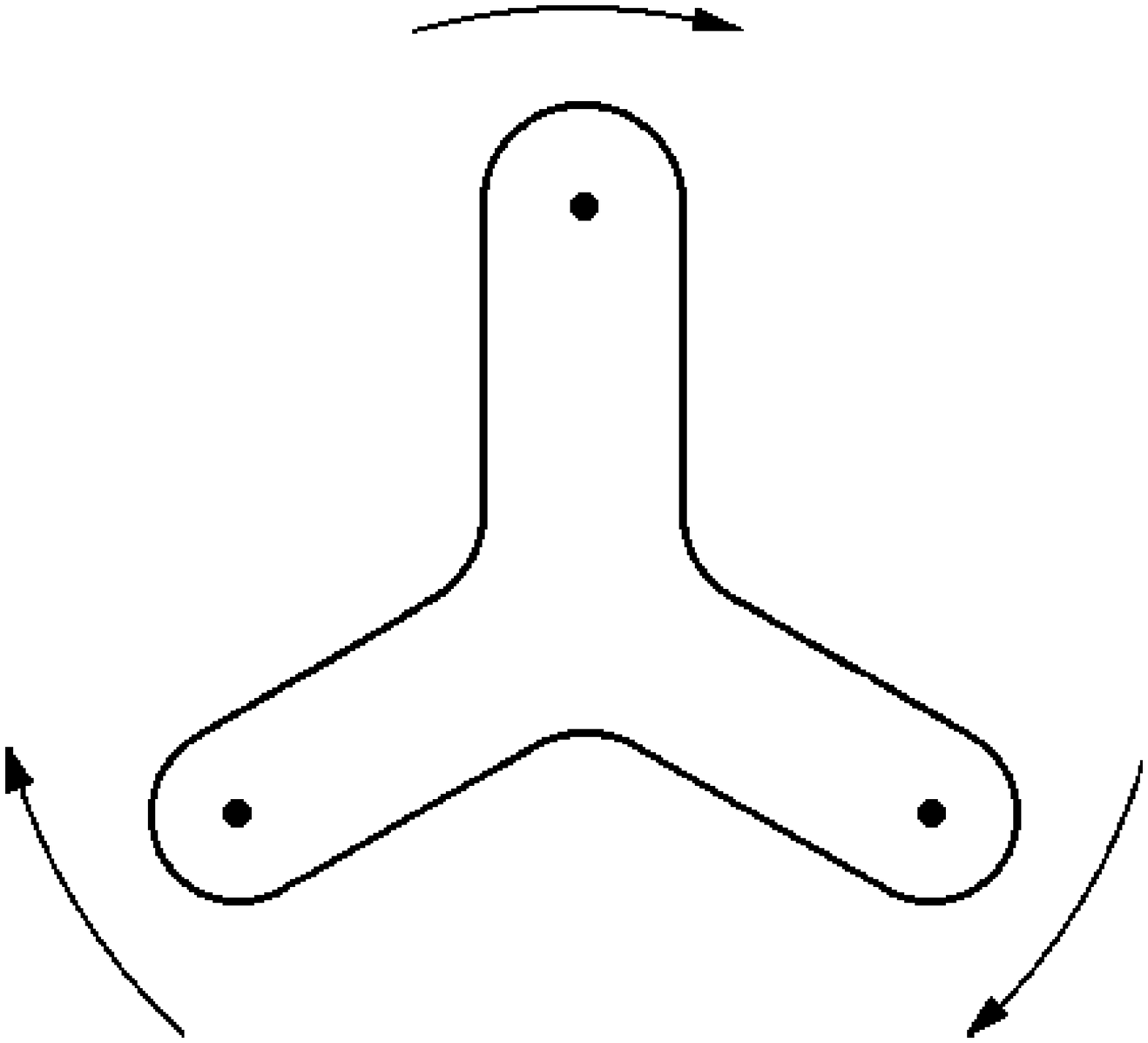}}
\end{center}
\caption{}
\end{figure}

%%%%%%%%%%%%%%%%%%%%%%%%%%%%%%%%%%%%%%%%%%%%%%%%%%%%%%%%%%%%%%%%%%%%%

\end{document}